# Terahertz Metamaterials with Semiconductor Split-Ring Resonators for Magnetostatic Tunability

**Jiaguang Han,[1] Akhlesh Lakhtakia,[2] and Cheng-Wei Qiu[3,*]**

[1]*Department of Physics, National University of Singapore, 2 Science Drive 3, Singapore 117542, Singapore*
*Email Address: phyhanj@nus.edu.sg*
[2]*Department of Physics, Indian Institute of Technology Kanpur, Kanpur 208016, India*
*and*
*Department of Engineering Science and Mechanics, Pennsylvania State University, University Park, PA 16802, USA (permanent address). Email Address: akhlesh@psu.edu*
[3] *Department of Electrical and Computer Engineering, National University of Singapore, 4 Engineering Drive 3, Singapore 117576, Singapore*
[*]*Corresponding author: eleqc@nus.edu.sg*

**Abstract:** We studied a metasurface constituted as a periodic array of semiconductor split-ring resonators. The resonance frequencies of the metasurface excited by normally incident light were found to be continuously tunable in the terahertz regime through an external magnetostatic field of suitable orientation. As such metasurfaces can be assembled into 3D metamaterials, the foregoing conclusion also applies to metamaterials comprising semiconductor split-ring resonators.

## 1. Introduction

Negatively refracting metamaterials have been a subject of recent enormous interest, in both the physics and the engineering communities, because they possess electromagnetic response characteristics that are not (at least, widely) displayed by natural materials [1,2,3]. Generally, these metamaterials are composite materials comprising metallic inclusions of specific shapes and sizes immersed in some homogeneous host medium, and convey the promise of a wide-ranging set of applications.

Metamaterial-based devices could come to include filters, modulators, amplifiers, transistors, and resonators, among others [4,5]. The usefulness of such a device could be extended tremendously if the metamaterial's response characteristics can be dynamically tuned. Tunability strategies examined thus far are electrical control [6,7,8], magnetostatic control [9,10], and optical pumping [11]. In all of cited publications, the inclusions are split-ring resonators (SRRs) made of some metal. Tunability is introduced by ensuring that another component (e.g., the substrate on which the SRRs are printed or another type of inclusions) is made of an electro-optic material, liquid crystal, ferrite, etc.

In this letter, we suggest a new type of tunable metamaterials. A metamaterial of this type comprises SRRs that are not metallic but, instead, are made of a material whose electromagnetic response properties can be tuned by an external agent. For the sake of illustration, we chose to focus on a tunable, terahertz, metamaterial that is controllable by an externally applied magnetostatic (or quasimagnetostatic) field.

## 2. Analysis and Numerical Results

The essence of the chosen metamaterial is a metasurface [12]: a planar array of semiconductor split-ring resonators (SRRs) periodically printed on an isotropic dielectric substrate. Figure 1a shows a single SRR with linear dimensions in the plane ranging from 2 μm to 36 μm, and with a thickness of 200 nm. The SRRs are printed, as shown in Fig. 1b, on a square lattice of period 60 μm. The SRRs are assumed to be made of a doped semiconductor, such as InAs, whose relative permittivity obeys the Drude model in the absence of an externally applied magnetostatic field [13,14]. As a typical isotropic dielectric substrate would only slightly shift the resonances but not affect other characteristics significantly [15], we set its electromagnetic properties to be that of free space (i.e., vacuum) in our numerical work.

Computer simulations of the spectral response of the chosen metasurface were performed using the commercial software CST Microwave Studio TM 2006B, which is a 3D full-wave solver that employs the finite integration technique. The planar SRR array was taken to be entirely surrounded by air and open boundary conditions were employed along the propagation direction. Without the applied magnetostatic field, the doped semiconductor's relative permittivity scalar $\varepsilon$ is given as a function of the angular frequency $\omega$ by

$$\varepsilon(\omega) = \varepsilon_\infty - \frac{\omega_p^2}{(\omega^2 + i\gamma\omega)}, \quad (1)$$

where $\varepsilon_\infty$ represents the high-frequency value; the plasma frequency $\omega_p = \sqrt{Ne^2/\varepsilon_0 m^*}$ depends on the carrier density *N*, the effective mass $m^*$, the electronic charge $e$, and the free-space permittivity $\varepsilon_0$; and $\gamma$ is the damping constant. The following parameters were used in our simulations [13,14]: $\varepsilon_\infty = 16.3$, $N = 1.04 \times 10^{-11}$ m$^{-3}$, $m^* = 0.004\, m_e$, and $\gamma/2\pi = 7.5 \times 10^{11}$ Hz, where $m_e$ is the electronic mass.

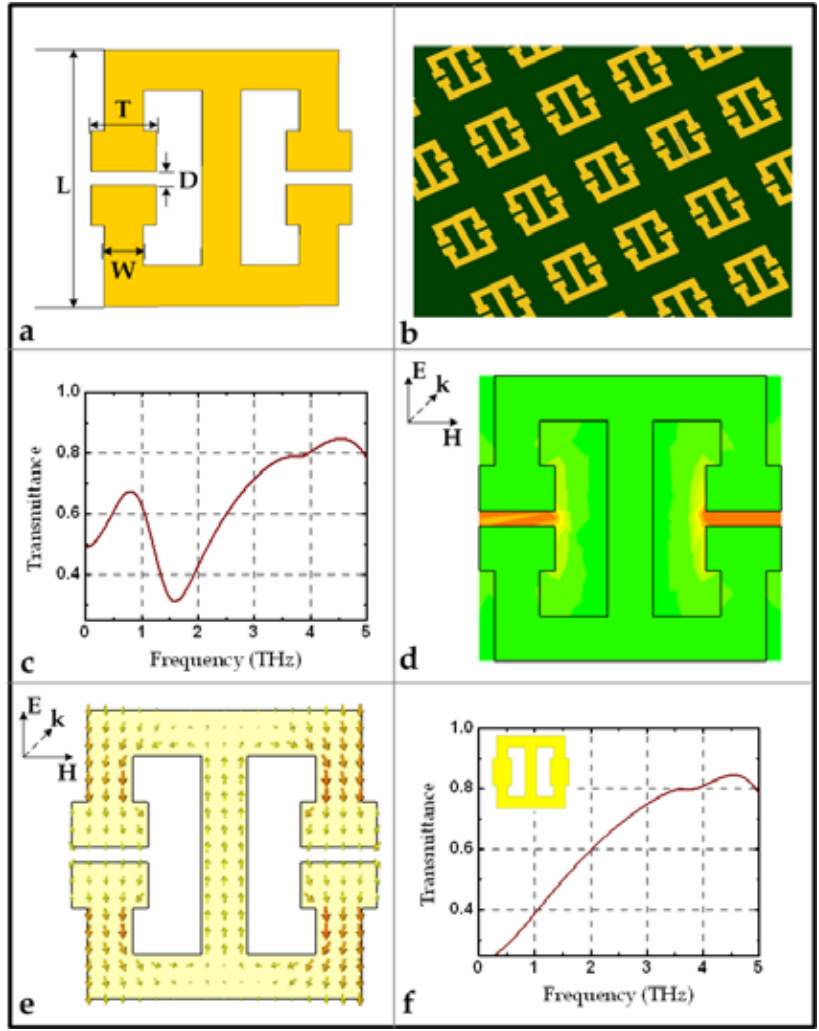

**FIG. 1**. (a) A single SRR with dimensions L = 36 μm, T = 10 μm, D = 2 μm and W = 6 μm. (b) The chosen metasurface, which is a square array of SRRs printed on a substrate with a period of 60 μm. (c) Transmittance spectrum of the chosen metasurface illuminated normally by a plane wave with electric field oriented perpendicular to the gaps in the SRR. (d) Map of the magnitude of the electric field in an SRR at 1.59 THz. (e) Map of the current density in an SRR at 1.59 THz. (f) The same as (c) but when both gaps are closed in every SRR

Let the chosen metasurface be oriented parallel the *xy* plane with the two gaps of every SRR aligned parallel to the *x* axis. The SRR gaps would resonate if the incident electric field were to be oriented along the *y* axis (i.e., perpendicular to the two gaps in each SRR).

We calculated the transmittance spectrum of the metasurface when illuminated by a plane wave with its wave vector oriented parallel to the *z* axis and its electric field oriented parallel to the *y* axis. Fig. 1c shows that a transmittance dip exists at 1.59 THz frequency. This dip can be attributed to a resonance of the SRRs: as shown in Fig. 1d, the electric field is concentrated in both gaps of every SRR at this frequency. Fig. 1e displays the vector plot of the current density at 1.59 THz. If both gaps of every SRR were closed, the transmittance dip at 1.59 THz would disappear, as can be gathered from the transmittance spectrum presented in Fig. 1f.

In order to obtain a tunable response from the chosen metasurface, we considered the application of an external magnetostatic field $\boldsymbol{B_0}$ to affect the resonance in the transmittance spectrum displayed in Fig. 1c. Generally, without the applied magnetostatic field, the dielectric response of a semiconductor is described by a scalar $\varepsilon(\omega)$, such as provided in Eq. (1). However, when $\boldsymbol{B_0}$ is applied, the scalar $\varepsilon(\omega)$ has to be replaced by the tensor $\hat{\varepsilon}(\omega)$.

In the Faraday configuration, $\boldsymbol{B_0}$ is aligned parallel to the wave vector of the incident plane wave (i.e., along the $z$ axis). With the assumption that $\boldsymbol{B_0}$ is spatially uniform, the nonvanishing components of $\hat{\varepsilon}(\omega)$ are [13]:

$$\varepsilon_{xx} = \varepsilon_{yy} = \varepsilon_\infty - \frac{\omega_p^2(\omega^2 + i\gamma\omega)}{(\omega^2 + i\gamma\omega)^2 - \omega^2\omega_c^2}, \qquad (2)$$

$$\varepsilon_{xy} = -\varepsilon_{yx} = i\frac{\omega\omega_c\omega_p^2}{(\omega^2 + i\gamma\omega)^2 - \omega^2\omega_c^2}, \qquad (3)$$

$$\varepsilon_{zz} = \varepsilon_\infty - \frac{\omega_p^2}{(\omega^2 + i\gamma\omega)}, \qquad (4)$$

where $\omega_c = eB_0/m^*$ is the cyclotron frequency. The off-diagonal components $\varepsilon_{xy}$ and $\varepsilon_{yx}$ are induced by the magnetostatic field and depend on its magnitude. Figure 2a contains the simulated transmittance spectra of the chosen metasurface for different values of the magnitude $B_0$ of $\boldsymbol{B_0}$. The transmittance dip redshifts gradually from 1.55 to 0.30 THz as $B_0$ is increased from 1.0 to 8.0 T, as also illustrated in Fig. 2b. Although the minimum transmittance increases as $B_0$ is increased, Figs. 2a and 2b clearly indicate that planar arrays of semiconducting SRRs are magnetically tunable in the terahertz regime.

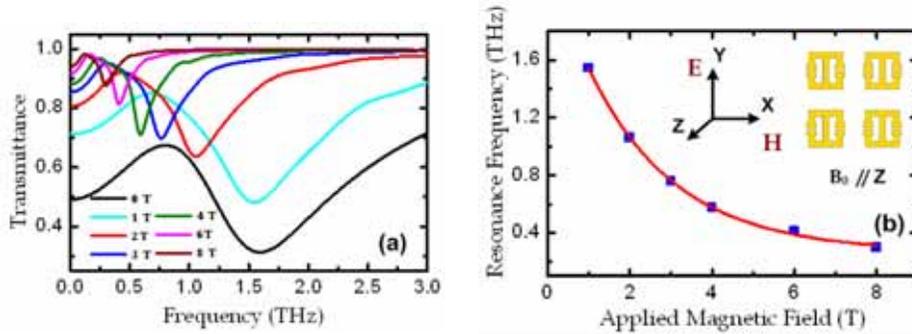

**FIG. 2**. Faraday configuration. (a) Transmittance spectra for various values of $B_0$, when $\boldsymbol{B_0}$ is aligned parallel to the wave vector of the incident plane wave and the electric field of the incident plane wave is aligned perpendicular to the gaps in each SRR. (b) Resonance frequency $\omega_0$ as a function of $B_0$, with the solid line showing the exponential fit $\omega_0 = \alpha + \beta\exp(-B_0/\chi)$, where $\alpha = 0.27$, $\beta = 2.04$, and $\chi = 2.11$.

Denoting the resonance angular frequency by $\omega_0$, we fitted the data in Fig. 2b to the formula $\omega_0 = 0.27 + 2.04\exp(-B_0/2.11)$. This formula indicates that the resonance frequency decays exponentially as the applied magnetostatic field increases in magnitude. One way to understand the foregoing trend is as follows. In the Faraday configuration, plane wave propagation in a semiconductor depends on [16], i.e., $\varepsilon_\pm = \varepsilon_{xx} \pm i\varepsilon_{xy} = \varepsilon_\infty - \frac{\omega_p^2}{\omega^2 + i\gamma\omega \mp \omega\omega_c}$. Let $\tilde{n}_\pm^2 = \varepsilon_\pm$, where $\tilde{n}_+$ and $\tilde{n}_-$ are the complex-valued refractive indexes for right- and left-circularly polarized plane waves. For the linearly polarized electromagnetic wave in our case, the response of substance can be captured by $\tilde{n} = \frac{1}{2}(\tilde{n}_+ + \tilde{n}_-)$ for different values of $B_0$. As shown in Fig. 3, the real part of $\tilde{n}$ decreases as $B_0$ increases. So the redshift of $\omega_0$ with increase of $B_0$ can be correlated with the concurrent decrease of $\tilde{n}$.

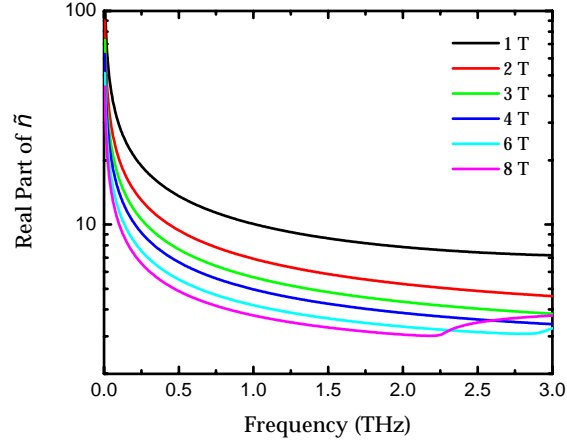

**FIG. 3**. Real part of $\tilde{n}$ as a function of the linear frequency $\omega/2\pi$ and the magnitude $B_0$ of the applied magnetostatic field in the Faraday configuration.

So far we have considered the Faraday configuration. In the Voigt configuration, the magnetostatic field is aligned perpendicular to the wave vector of the incident plane wave. There are two cases, depending on the direction of the incident electric field relative to $\boldsymbol{B_0}$. Suppose first that $\boldsymbol{B_0}$ is also aligned perpendicular to the electric field of the incident plane wave (i.e., $\boldsymbol{B_0}$ is oriented along the $x$ axis). The nonvanishing components of $\hat{\varepsilon}(\omega)$ now are [13]

$$\varepsilon_{yy} = \varepsilon_{zz} = \varepsilon_\infty - \frac{\omega_p^2(\omega^2 + i\gamma\omega)}{(\omega^2 + i\gamma\omega)^2 - \omega^2\omega_c^2}, \qquad (5)$$

$$\varepsilon_{yz} = -\varepsilon_{zy} = i\frac{\omega\omega_c\omega_p^2}{(\omega^2 + i\gamma\omega)^2 - \omega^2\omega_c^2}, \qquad (6)$$

$$\varepsilon_{xx} = \varepsilon_\infty - \frac{\omega_p^2}{(\omega^2 + i\gamma\omega)}. \qquad (7)$$

Figure 4a shows that the transmittance spectra in the first Voigt configuration are qualitatively similar to that for the Faraday configuration in Fig. 2a. The resonance frequency exponentially redshifts as $B_0$ increases, and can be fitted to $\omega_0 = 0.27 + 1.64\exp(-B_0/3.03)$; see Fig. 4b. Furthermore, in this configuration, the electromagnetic response of the semiconductor is related to $\tilde{n}_\perp = (\varepsilon_{yy} + \varepsilon_{yz}^2/\varepsilon_{yy})^{1/2}$. Figure 5 shows that real part of $\tilde{n}_\perp$ decreases with increasing $B_0$, which implies the same correlation with the resonance frequency as deduced for the Faraday configuration.

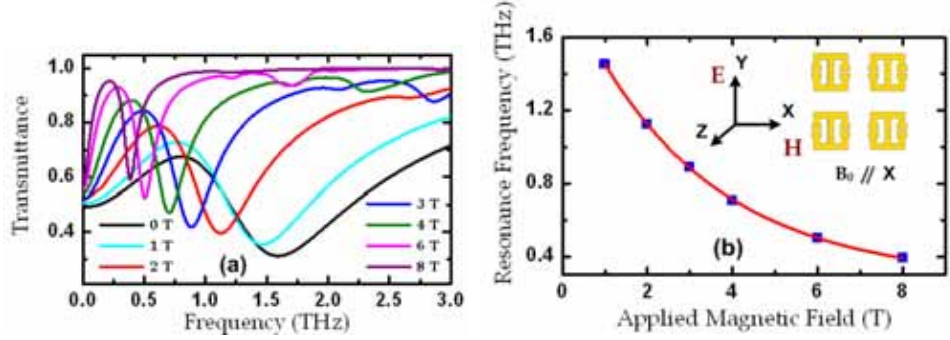

**FIG. 4**. First Voigt configuration. (a) Transmittance spectra for various values of $B_0$, when $\boldsymbol{B_0}$ is aligned parallel to the gaps in each SRR and the electric field of the incident plane wave is aligned perpendicular to the gaps in each SRR. (b) Resonance frequency $\omega_0$ as a function of $B_0$, with the solid line showing the exponential fit $\omega_0 = \alpha + \beta \exp(-B_0/\chi)$, where $\alpha = 0.27$, $\beta = 1.64$, and $\chi = 3.03$.

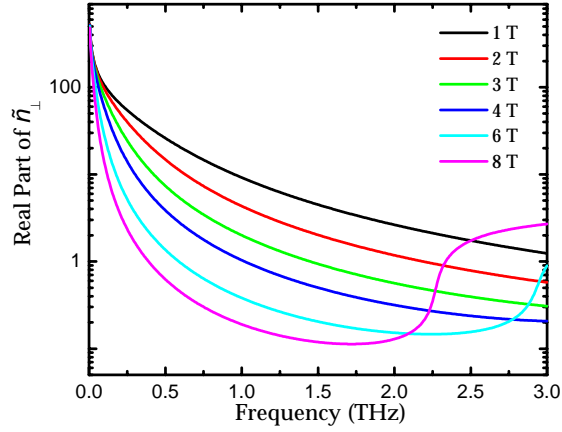

**FIG. 5**. Real part of $\tilde{n}_\perp$ as a function of the linear frequency $\omega/2\pi$ and the magnitude $B_0$ of the applied magnetostatic field in the first Voigt configuration.

In the second Voigt configuration, $\boldsymbol{B_0}$ is aligned parallel to the electric field of the incident plane wave (i.e., $\boldsymbol{B_0}$ is oriented along the $y$ axis), so that the nonvanishing components of $\hat{\varepsilon}(\omega)$ now are [13]

$$\varepsilon_{xx} = \varepsilon_{zz} = \varepsilon_\infty - \frac{\omega_p^2(\omega^2 + i\gamma\omega)}{(\omega^2 + i\gamma\omega)^2 - \omega^2\omega_c^2}, \quad (8)$$

$$\varepsilon_{zx} = -\varepsilon_{xz} = i\frac{\omega\omega_c\omega_p^2}{(\omega^2 + i\gamma\omega)^2 - \omega^2\omega_c^2}, \quad (9)$$

$$\varepsilon_{yy} = \varepsilon_\infty - \frac{\omega_p^2}{(\omega^2 + i\gamma\omega)}. \quad (10)$$

We found that there is no remarkable effect of the applied magnetostatic field on the resonance frequency of the chosen metasurface; furthermore, the electromagnetic response of the semiconductor is related to $\tilde{n}_\parallel = \varepsilon_{yy}^{1/2}$, which does not depend on $B_0$.

## 3. Conclusion

To conclude, our computer simulations have shown that the terahertz resonant response of a metasurface constituted by a planar array of semiconducting split-ring resonators can be tuned by the application of an externally applied magnetostatic field of appropriate orientation and magnitude. As such metasurfaces can be assembled into 3D metamaterials, the foregoing conclusion applies to metamaterials comprising semiconducting SRRs as well. SRRs made of other tunable materials, such as electro-optic materials, could also assist in the realization of tunable terahertz metamaterials.

## 4. Acknowledgements

J.H. acknowledges financial support from the MOE Academic Research Fund of Singapore and fruitful discussions with J. Q. Gu. A.L. thanks the Department of Physics, IIT Kanpur for its hospitality.